\documentclass[prd,preprint,nofootinbib,amsmath,amssymb]{revtex4-1}
\usepackage{amsmath, amssymb, mathtools}
\usepackage{bm}
\usepackage{mathrsfs}
\usepackage[colorlinks=true,linkcolor=blue,citecolor=blue]{hyperref}
\usepackage{cleveref}

\newcommand{\be}{\begin{equation}}
\newcommand{\ee}{\end{equation}}

\usepackage{color}

\usepackage{breqn}
\makeatletter
\let\cat@comma@active\@empty
\makeatother

\begin{document}
	
\title{\boldmath Near--$AdS_2$ perturbations and the connection \\ with near-extreme Reissner-Nordstrom}
	
\author{Achilleas P. Porfyriadis}
\affiliation{Department of Physics, UCSB, Santa Barbara, CA 93106, USA}
	
	
\begin{abstract}
The geometry very near the horizon of a near-extreme Reissner-Nordstrom black hole is described by the direct product of a near--$AdS_2$ spacetime with a two-sphere. While near--$AdS_2$ is locally diffeomorphic to $AdS_2$ the two connect differently with the asymptotically flat part of the geometry of (near-)extreme Reissner-Nordstrom. In previous work, we solved analytically the coupled gravitational and electromagnetic perturbation equations of $AdS_2\times S^2$ and the associated connection problem with extreme Reissner-Nordstrom. In this paper, we give the solution for perturbations of near--$AdS_2\times S^2$ and make the connection with near-extreme Reissner-Nordstrom. Our results here may also be thought of as computing the classical scattering matrix for gravitational and electromagnetic waves which probe the region very near the horizon of a highly charged spherically symmetric black hole.
\end{abstract}
	
\maketitle

	
\section{Introduction}

The direct products of a two-sphere with $AdS_2$ or near--$AdS_2$ are exact solutions of four-dimensional Einstein-Maxwell theory without a cosmological constant. In \cite{paper1} we analytically solved the coupled gravitational and electromagnetic perturbation equations of $AdS_2\times S^2$ in this theory. While $AdS_2$ and near--$AdS_2$ are locally diffeomorphic, the solutions to the corresponding perturbation equations give rise to distinct answers for the so-called connection problem. This is the problem of extending anti-de Sitter solutions away from the near-horizon region of (near-)extreme black holes and connecting them with solutions in the far asymptotically flat region. In \cite{paper1} we solved the connection problem for the gravitational and electromagnetic perturbations of $AdS_2\times S^2$. Namely, we produced the $AdS_2\times S^2$ perturbation equations as an appropriate near-horizon approximation of the corresponding equations for the extreme Reissner-Nordstrom black hole and then, using matched asymptotic expansions, we analytically extended the $AdS_2\times S^2$ solutions away from the near-horizon region connecting them with solutions in the far asymptotically flat region. In this paper, we solve the connection problem for perturbations of near--$AdS_2$, making the connection with near-extreme Reissner-Nordstrom.

From the high-energy theory viewpoint, the motivation for studying the connection problem stems from a desire to pave the way for transferring results from (near--)$AdS_2$ holography to the realm of four-dimensional
classical and quantum gravity in asymptotically flat spacetimes. 
This viewpoint is explained in more detail in \cite{paper1}, where one may also find a more extensive list of references. 
On the other hand, it is worthwhile to emphasize that the results contained here --and in \cite{paper1}-- are of interest from a purely classical gravitational physics viewpoint as well. In particular, here we present a new take on the venerable problem of Reissner-Nordstrom perturbations whereby the perturbation equations are reduced to a single fourth-order radial differential equation on the metric perturbation that measures the deviation in the size of the background two-sphere. Moreover, for a background that is nearly extremal and at low energies we solve this equation analytically using the method of matched asymptotic expansions.

In Section \ref{sec: review NERN, AdS2xS2, NAdS2xS2, RWZ} we set notation, briefly reviewing the $AdS_2\times S^2$ and near--$AdS_2\times S^2$ throat geometries of near-extreme Reissner-Nordstrom (NERN), as well as the Regge-Wheeler-Zerilli gauge for the pertubation equations.  
Section \ref{sec: reduction of perturbation eqs} offers a reduction of the perturbation equations for a general Reissner-Nordstrom black hole to a single fourth-order radial differential equation. This is an alternative to the well known reductions to two second-order equations.
In Section \ref{sec: main} we obtain the exact analytic answer for near--$AdS_2\times S^2$ perturbations and solve the connection problem using matched asymptotic expansions in NERN.
Section \ref{sec: in and up solutions} specializes the connection formulas to a basis of purely ingoing near the horizon and purely outgoing near infinity solutions.
	
\section{\boldmath Near-Extreme Reissner-Nordstrom, $AdS_2\times S^2$, and $NAdS_2\times S^2$}\label{sec: review NERN, AdS2xS2, NAdS2xS2, RWZ}

The general Reissner-Nordstrom black hole of mass $M$ and charge $Q$ is given by:\footnote{$G=c=1$}
\be\label{RN hatted metric}
ds^2=-\left(1-{2M\over\hat{r}}+{Q^2\over\hat{r}^2}\right)d\hat{t}^2+\left(1-{2M\over\hat{r}}+{Q^2\over\hat{r}^2}\right)^{-1}d\hat{r}^2+\hat{r}^2d\Omega^2\,,\qquad \hat{A}_{\hat{t}}=-{Q\over \hat{r}}\,.
\ee
The near-extreme Reissner-Nordstrom (NERN) is characterized by:
\be
\kappa\equiv\sqrt{1-{Q^2\over M^2}}\ll 1\,.
\ee
NERN has a long throat to the horizon and we can derive regular throat geometries as follows \cite{Maldacena:1998uz,Spradlin:1999bn}. Begin by making the coordinate and gauge transformation,
\be
r=\frac{\hat{r}-r_+}{r_+}\,,\quad t=\frac{\hat{t}}{M}\,,\qquad A=\hat{A}+d\hat{t}\,,
\ee
with $r_\pm=M(1\pm\kappa)$, to obtain:
\begin{align}\label{NERN lambda metric}
\begin{aligned}
{1\over M^2} ds^2 &= -\frac{r(r+2\kappa+\kappa r)}{(1+\kappa)(1+r)^2} dt^2 +\frac{(1+\kappa)^3(1+r)^2}{r(r+2\kappa+\kappa r)} dr^2+(1+\kappa)^2(1+r)^2d\Omega^2\,,\\
A_t&=M\bigg(1-\sqrt{1-\kappa\over 1+\kappa}{1\over 1+r}\bigg)\,.
\end{aligned}
\end{align}
Taking the limit $r\sim\kappa\ll 1$ we get:
\be\label{NAdS2xS2}
{1\over M^2} ds^2= -r(r+2\kappa) dt^2 + \frac{dr^2}{r(r+2\kappa)}+d\Omega^2\,,\qquad A_t=M(r+\kappa)\,,
\ee
while the limit $r\sim\kappa^p\ll 1$, with $0<p<1$, produces:
\be\label{AdS2xS2}
{1\over M^2} ds^2= -r^2 dt^2 + \frac{dr^2}{r^2}+d\Omega^2\,,\qquad A_t=M r\,.
\ee
Both \eqref{NAdS2xS2} and \eqref{AdS2xS2} solve the Einstein-Maxwell equations on their own. We will refer to the metric in \eqref{NAdS2xS2} as $NAdS_2\times S^2$ and the metric in \eqref{AdS2xS2} simply as $AdS_2\times S^2$. Here, ``$NAdS$'' stands for ``nearly anti-de Sitter.'' Locally, $NAdS_2$ and $AdS_2$ are diffeomorphic and on a Penrose diagram of the throat one finds that \eqref{NAdS2xS2} covers a Rindler patch while \eqref{AdS2xS2} covers a Poincare patch.\footnote{For readers who are familiar with the throat geometries of near-extreme Kerr we note that $NAdS_2\times S^2$ is to $AdS_2\times S^2$ what near-NHEK \cite{Bredberg:2009pv} is to NHEK \cite{Bardeen:1999px}.}
From now on we set $M=1$.

In this paper, as in \cite{paper1}, we will solve the even parity sector of the linearized Einstein-Maxwell equations for all $l\geq 2$ modes in the Regge-Wheeler-Zerilli gauge \cite{Regge:1957td, Zerilli:1974ai}:
\begin{align}
h_{\mu\nu}=&
\begin{pmatrix}
-Y(r)  & X(r)	& 0 	& 0 \\
& V(r)	& 0 	& 0 \\
& 	 	& K(r) 	& 0 \\
&  		&  		& K(r)\sin^2\theta
\end{pmatrix}
e^{i\omega t}\, Y_{l,0}(\theta,\phi) \,,\label{metric pertn ansatz}\\
a_\mu=&
\begin{pmatrix}
\chi(r)  & \psi(r)	& 0 	& 0 
\end{pmatrix}	
e^{i\omega t}\, Y_{l,0}(\theta,\phi)\,. \label{gauge field pertn ansatz}
\end{align}
Here $h_{\mu\nu}\,, a_\mu$ denote the perturbations to the background metric and gauge field, respectively.
Note that we are considering only the even parity sector because we are mainly interested in perturbations to the $NAdS_2$ part of the near-horizon metric. Also, we have used the spherical symmetry of the background to set $m=0$. Finally, we remind the reader that the $l=0,1$ cases are special and need to be treated separately.\footnote{The technical reason for this are the special facts that $\partial_\theta Y_{0,0}=0$ and $\partial^2_\theta Y_{1,0}=-Y_{1,0}$ which imply that certain components of the linearized Einstein-Maxwell equations vanish identically for the $l=0,1$ modes.}

\section{Reduction of the perturbation equations}\label{sec: reduction of perturbation eqs}

Using the ansatz (\ref{metric pertn ansatz}--\ref{gauge field pertn ansatz}) and following the reduction path described in \cite{paper1}, we may reduce the linearized Einstein-Maxwell equations on the background of the general Reissner-Nordstrom black hole \eqref{NERN lambda metric}, to a single fourth-order differential equation for $K$,
\be\label{Keqn}
a_4(r)K''''+a_3(r)K'''+a_2(r)K''+a_1(r)K'+a_0(r)K=0\,,
\ee
with the expressions for the coefficients $a_i(r)$ given in App.~\ref{app: a's}.
The remaining components of the perturbation ansatz (\ref{metric pertn ansatz}--\ref{gauge field pertn ansatz}) are then given by:
\begin{dgroup*}
		
	\begin{dmath}\label{Y soln}
	Y=
	-\frac{1}{(1+\kappa)^4 (1+r)^6 \left(\lambda^2+4 (1+\kappa)^2 (1+r)^2 \omega^2\right)}
	\\
	\times\left[\lambda r^2 (1+r)^2 (r+2\kappa+\kappa r)^2 K'' 
	-2 r (1+r) (r+2 \kappa +\kappa r) \left(\lambda r (r+2\kappa+\kappa r) -2(1+\kappa)^3 (1+r)^4 \omega^2 \right) K' + 
	\left(2 \lambda r^2 (r+2\kappa+\kappa r)^2-(1+\kappa)^3 (1+r)^5 \omega^2 \left(\lambda (1+\kappa) (1+r)+4 (r+\kappa +\kappa  r)\right)\right)K  \right]
	\end{dmath}
		
	\begin{dmath}\label{V soln}
		V=-\frac{(1+\kappa)^4(1+r)^4}{r^2(r+2\kappa+\kappa r)^2} Y
	\end{dmath}
	
	\begin{dmath}\label{X soln}
		X=
		\frac{2i\omega\left((1+\kappa) (1+r) \left(r(r+2\kappa+\kappa r) K' -(r+\kappa+\kappa r)K \right)-r^2 (r+2\kappa +\kappa  r)^2 V\right)}
		{\lambda (1+\kappa) r (1+r) (r+2 \kappa +\kappa r)}
	\end{dmath}
	
	\begin{dmath}\label{chi soln}
		\chi=
		\frac{1}{4\sqrt{1-\kappa^2}(1+\kappa)^3 r(1+r)^3 (r+2\kappa+\kappa r)} \times 
		\left[
		(1+\kappa) r^2 (1+r) (r+2 \kappa +\kappa r)^2 K'
		-2 (1+\kappa) r^2 (r+2 \kappa +\kappa r)^2 K
		+(1+\kappa)^4 r(1+r)^5 (r+2 \kappa +\kappa r) Y'
		-(1+\kappa)^4 (1+r)^5 (r+\kappa +\kappa r) Y
		-r^2 (1+r) (r+\kappa+\kappa r) (r+2 \kappa +\kappa r)^2 V
		+i\omega (1+\kappa)^4 r (1+r)^5 (r+2 \kappa +\kappa r)X
		\right]
	\end{dmath}
	
	\begin{dmath}\label{psi soln}
		\psi=
		\frac{1+r}{4\sqrt{(1-\kappa)/(1+\kappa )} \, r (r+2 \kappa +\kappa  r)} \times 
		\left(i\omega (1+\kappa) (1+r) K
		+i\omega r (1+r) (r+2 \kappa +\kappa r) V
		-2 (r+\kappa) X
		-r (1+r) (r+2 \kappa +\kappa r)X'
		\right)
	\end{dmath}
	
\end{dgroup*}
where $\lambda\equiv l(l+1)$.

Note that the above reduction of the perturbation equations does not assume near-extremality and is valid for the general Reissner-Nordstrom black hole.
It is an alternative to the well known reductions to two second-order equations and contains the same information.

For readers' convenience a \emph{Mathematica} notebook with Eqs.~(\ref{Y soln}--\ref{psi soln}) and the coefficients $a_i(r)$ from App.~\ref{app: a's} is included with this paper.

\section{\boldmath $NAdS_2\times S^2$ answers and the connection problem solved}\label{sec: main}

Consider the scattering of gravitational and electromagnetic waves in the full NERN spacetime. Such waves will survive the near-horizon limit and probe the $NAdS_2\times S^2$ region of the spacetime if and only if their energy is low:
\be
\omega\ll 1\,.
\ee
In this regime, to leading order in $\kappa$ and $\omega$, we may solve the scattering problem analytically using the method of matched asymptotic expansions.
To this end, as in \cite{paper1}, we change variables in Eq.~\eqref{Keqn} according to
\be \label{KtoH}
K(r)=\left(\frac{r}{1+r}\right)^2 H(r) \,,
\ee
and solve the resulting equation for $H$,
\be\label{Heqn}
b_4(r)H''''+b_3(r)H'''+b_2(r)H''+b_1(r)H'+b_0(r)H=0\,,
\ee
by dividing the spacetime into three regions:
\begin{align}
&\textrm{Near:}\qquad\qquad\qquad\quad\quad\,\,\,\,\, r\ll 1 \label{near region}\\
&\textrm{Static:}\qquad\, \textrm{max}(\kappa,\omega)\ll \,r\ll 1/\omega\label{static region}\\
&\textrm{Far:}\qquad\qquad\qquad\quad 1\ll\,r \label{far region}
\end{align}
Note that compared to \cite{paper1} the Static region here is defined by imposing the additional condition $r\gg\kappa$. Nevertheless, for NERN, this Static region still overlaps with both the Near and the Far regions and may therefore again be used to connect the Near and Far solutions.

\subsection{Near region} 
In the Near region, which corresponds to $NAdS_2\times S^2$, Eq.~\eqref{Heqn} reduces to
\begin{align*}
&b_4^{n}(r)=r^4 (r+2 \kappa)^4\\
&b_3^{n}(r)=8 r^3 (r+2 \kappa )^3 (2 r+3 \kappa)\\
&b_2^{n}(r)=-2 r^2 (r+2 \kappa )^2 \left(\left(l^2+l-36\right) r^2
+2 \left(l^2+l-54\right)\kappa r -76 \kappa ^2 -\omega ^2  \right)\\
&b_1^{n}(r)=-4 r (r+2 \kappa)^2 \left(3 \left(l^2+l-8\right) r^2+5 (l-3) (l+4)\kappa r -32 \kappa^2-2 \omega ^2\right)\\
&b_0^{n}(r)= (r+2 \kappa)^2 \left((l-3) (l-1) (l+2) (l+4) r^2 -16 \left(l^2+l-3\right)\kappa r +16 \kappa ^2\right) \notag\\
&\qquad\quad-2\left(\left(l^2+l-2\right) r^2+ 
2\left(l^2+l-4\right)\kappa r  -10 \kappa ^2 \right)\omega^2  +\omega ^4
\end{align*}
and the solution is given by
\begin{align}\label{Hnear soln}
H^{n}(r)=
C_1^n &\, r^{-2-\frac{i \omega }{2 \kappa }} 
	\left({r\over 2 \kappa}+1\right)^{\frac{i \omega }{2 \kappa }} \, _2F_1\left(-1-l,l+2;1-\frac{i\omega }{\kappa };-\frac{r}{2 \kappa }\right) \notag\\+
C_2^n &\, r^{-2+\frac{i \omega }{2 \kappa }} 
	\left({r\over 2 \kappa}+1\right)^{-\frac{i \omega }{2 \kappa }} \, _2F_1\left(-1-l,l+2;1+\frac{i \omega }{\kappa };-\frac{r}{2 \kappa }\right) \notag\\+
C_3^n &\, r^{-2-\frac{i \omega }{2 \kappa }} 
	\left({r\over 2 \kappa}+1\right)^{\frac{i \omega }{2 \kappa }} \, _2F_1\left(1-l,l;1-\frac{i	\omega }{\kappa };-\frac{r}{2 \kappa }\right)\notag\\+
C_4^n &\, r^{-2+\frac{i \omega }{2 \kappa }} 
	\left({r\over 2 \kappa}+1\right)^{-\frac{i \omega }{2 \kappa }} \, _2F_1\left(1-l,l;1+\frac{i \omega }{\kappa };-\frac{r}{2 \kappa }\right) \,.
\end{align}
Here $_2F_1(a,b;c;z)$ is the Gauss hypergeometric function.\footnote{Recall that for $m=0,1,2,\ldots$ and $c\neq 0,-1,-2,\ldots$ we have the following polynomial:
\[
_2F_1(-m,b;c;z)=\sum_{k=0}^{m}(-1)^k\binom{m}{k}\frac{(b)_k}{(c)_k}z^k
\]	
with the (rising) Pochhammer symbol defined as $(a)_k=a(a+1)(a+2)\cdots(a+k-1)$.
}

This Near solution is an exact solution with respect to the $NAdS_2\times S^2$ background \eqref{NAdS2xS2}. Specifically, using the perturbation ansatz (\ref{metric pertn ansatz}--\ref{gauge field pertn ansatz}) on the background \eqref{NAdS2xS2}, Eq.~\eqref{Hnear soln} gives the exact solution of the linearized Einstein-Maxwell equations, provided that Eqs.~(\ref{Y soln}--\ref{psi soln}) are replaced by
\begin{align*}
&Y=-\frac{r^2 (r+2 \kappa )^2 K''-\omega ^2 K}{l (l+1)}\,, 
V=-\frac{1}{r^2 (r+2 \kappa )^2}Y\,,
X=-\frac{2 i \omega  \left((r+\kappa ) K-r (r+2 \kappa ) K'\right)}{l (l+1) r (r+2 \kappa )}\,,\\
&\chi=\frac{r^2 (r+2 \kappa )^2 K'-(r+\kappa ) Y+r (r+2 \kappa ) Y'
	-r^2 (r+2 \kappa )^2 (r+\kappa ) V+i \omega  r (r+2 \kappa) X} {4r(r+2\kappa)}\,,\\
&\psi=\frac{i\omega K+i\omega r(r+2 \kappa)V-2(r+\kappa)X-r(r+2\kappa)X'}
	{4r(r+2 \kappa)}\,,
\end{align*}
and Eq.~\eqref{KtoH} is replaced by $K(r)=r^2 H(r)$. 
Note that because  $NAdS_2$ is locally diffeomorphic to $AdS_2$ one could have obtained this solution by starting from the solution for $AdS_2\times S^2$ found in \cite{paper1}, applying the appropriate diffeomorphism (given \emph{e.g.} in \cite{Spradlin:1999bn}), and then adjusting the gauge to return to the Regge-Wheeler-Zerilli one.

\subsection{Static and Far regions}

In the Static and Far regions, we find that Eq.~\eqref{Heqn} reduces to the same equations as in the case of exactly extreme Reissner-Nordstrom studied in \cite{paper1}. Intuitively, the reason for this is the fact that for $r\gg\kappa$ we are in a region far enough from the horizon that we can no longer discern the black hole's small deviation from extremality.
Note, however, that this intuitive explanation has a precise realization for Eq.~\eqref{Heqn} only because we have successfully set up the problem in the correct variables such that the Static and Far approximations \eqref{static region} and \eqref{far region}, respectively, when applied on the expressions in App.~\ref{app: a's} realize this intuition.
For convenience, here we recall the Far solution \cite{paper1}:
\begin{align}\label{Hfar soln}
H^{f}(r)=
C_1^f &\,  r \omega  
\left(
(l+1) (l+2) r \omega  h_{l+2}^{(1)}(r \omega )
-\left((l+1) (l+2) (2 l+3) -2 r^2 \omega^2\right) h_{l+1}^{(1)}(r \omega)
\right) \notag\\+
C_2^f &\, r \omega  
\left(
(l+1) (l+2) r \omega  h_{l+2}^{(2)}(r \omega )
-\left( (l+1) (l+2) (2 l+3) -2 r^2 \omega^2\right) h_{l+1}^{(2)}(r \omega)
\right)\notag\\+
C_3^f &\, \left(
r \omega  \left(l (l+1)^3 (l+2)-2 \left(l^2+l+2\right) r^2 \omega ^2\right) h_{l+2}^{(1)}(r \omega )\right.\notag\\
&\quad\left.-(l+2) \left(l (l+1)^3 (2 l+3)-(l+3)
\left(l^2+l+2\right) r^2 \omega ^2\right) h_{l+1}^{(1)}(r \omega )
\right)\notag\\+
C_4^f &\, \left(r \omega  \left(l (l+1)^3 (l+2)-2 \left(l^2+l+2\right) r^2 \omega
^2\right) h_{l+2}^{(2)}(r \omega )\right.\notag\\
&\quad\left.-(l+2) \left(l (l+1)^3 (2 l+3)-(l+3) \left(l^2+l+2\right) r^2 \omega ^2\right) h_{l+1}^{(2)}(r \omega )\right)\,.
\end{align}
where $h_{m}^{(1)},h_{m}^{(2)}$ are the spherical Hankel functions of the first and second kind, respectively.

\subsection{Overlap regions, matching, and the solution to the connection problem}

The overlap regions are given by
\begin{align} \label{overlap regions}
\textrm{Near-Static overlap:}\qquad\quad \,\textrm{max}(\kappa,\omega)&\ll \,r\ll 1 \\
\textrm{Far-Static overlap:}\qquad\qquad\qquad\quad \,1&\ll \,r\ll 1/\omega 
\end{align}
and the corresponding equations and solutions are again exactly as in the extremal case \cite{paper1}.

Matching the Near solution \eqref{Hnear soln} to the Static solution found in \cite{paper1} and using the matching of the Static and Far solutions already performed in \cite{paper1} we obtain the desired answer to the connection problem in NERN, \emph{i.e.} the following linear relation between the $C_i^f$'s and $C_i^n$'s:
\begin{align}
&\begin{aligned}
C_{12}^{n+}
=\,& \frac{i}{\pi}\frac{(l+1) (l+2) }{ 2 l-1}  2^{-2 l-2} \,
\Gamma \left(-l-\tfrac{1}{2}\right)^2 
\left((25l-8)\omega \,C_{12}^{f+}-3 l(l+1)^2 (3 l-1) \,C_{34}^{f+} \right)\times \\ 
&\times (2\kappa)^{i\omega/2\kappa}\kappa^{l+2}\omega^l ({-i\omega/\kappa})_{l+2}\notag
\end{aligned} 
\\
&\begin{aligned} \label{scattering eqns soln}
C_{12}^{n-}
=\,& -\frac{i}{\pi} (l-1) l (l+1) (2l+1) 2^{2l} \, 
\Gamma \left(l+\tfrac{1}{2}\right)^2 
\left(3\omega \,C_{12}^{f-} + l^2 (l+1)\, C_{34}^{f-} \right) \times \\
&\times \frac{(2\kappa)^{i\omega/2\kappa}}{\kappa^{l+2}\omega^l ({i\omega/\kappa})_{l+2}}
\\
C_{34}^{n+}
=\,& \frac{i}{\pi}  l(l+1) (l+2) (2l+1)  2^{-2l-2} \,
\Gamma \left(-l-\tfrac{1}{2}\right)^2 
\left(3 \omega \,C_{12}^{f+} - l (l+1)^2 \,C_{34}^{f+}\right) \times \\
&\times (2\kappa)^{i\omega/2\kappa}\kappa^{l}\omega^l ({-i\omega/\kappa})_{l}
\end{aligned}
\\
&\begin{aligned}
C_{34}^{n-}
=\,&-\frac{i}{\pi}\frac{ (l-1) l }{2 l+3} 2^{2l} \,
\Gamma \left(l+\tfrac{1}{2}\right)^2 
\left((25l+33)\omega \,C_{12}^{f-} +3 l^2(l+1)(3l+4) \,C_{34}^{f-}\right)\times \\ 
&\times \frac{(2\kappa)^{i\omega/2\kappa}}{\kappa^{l}\omega^l ({i\omega/\kappa})_{l}}\notag
\end{aligned}
\end{align}
where we have defined
\begin{align}
\begin{aligned}
C_{12}^{n\pm}&=C_{1}^{n}\pm \alpha \, C_{2}^{n}\,, & C_{34}^{n\pm}&=C_{3}^{n}\pm \beta \, C_{4}^{n}\,,\\
C_{12}^{f\pm}&=C_{1}^{f}\pm C_{2}^{f}\,, & 
C_{34}^{f\pm}&=C_{3}^{f}\pm C_{4}^{f}\,,
\end{aligned}
\end{align}
with $\alpha=-(2 \kappa )^{{i \omega}/{\kappa} } 
{\left(-{i \omega }/{\kappa}\right)_{l+2}}/{\left({i \omega }/{\kappa}\right)_{l+2}}$ and  $\beta=-(2 \kappa )^{{i \omega }/{\kappa }} 
{\left(-{i \omega }/{\kappa}\right)_{l}}/{\left({i \omega }/{\kappa}\right)_{l}}\,.$
This completes the solution to the connection problem for $NAdS_2\times S^2$. 

The connection formulas \eqref{scattering eqns soln} may also be thought of as computing the classical scattering matrix for gravitational and electromagnetic waves which probe the near-horizon $NAdS_2\times S^2$ region of a NERN black hole.
Notice that while the Far solution \eqref{Hfar soln} is identical for the near-extreme black hole in this paper and the exactly extreme one in \cite{paper1}, the connection formulas with $NAdS_2\times S^2$ solutions here lead to the markedly distinct results of this paper compared to the results obtained by making the connection with $AdS_2\times S^2$ solutions in \cite{paper1}.

\section{A useful basis of solutions}\label{sec: in and up solutions}

Consider the basis of solutions of Eq.~\eqref{Heqn} that consists of two solutions $H_1, H_2$ which are purely ingoing near the horizon and two solutions $H_3, H_4$ which are purely outgoing near infinity.\footnote{The caveats explained in \cite{paper1} regarding the boundary conditions obeyed by the analogous basis in the case of extreme Reissner-Nordstrom apply here as well.} This basis may be defined as follows:
\begin{align}
H_1 \quad \textrm{is the solution with:}\quad C^n_2=1\,,C^n_1=C^n_3=C^n_4=0\,,\\
H_2\quad \textrm{is the solution with:}\quad C^n_4=1\,,C^n_1=C^n_3=C^n_2=0\,,\\
H_3 \quad \textrm{is the solution with:}\quad C^f_2=1\,,C^f_1=C^f_3=C^f_4=0\,,\\
H_4\quad \textrm{is the solution with:}\quad C^f_4=1\,,C^f_1=C^f_3=C^f_2=0\,.
\end{align}
From the connection formulas \eqref{scattering eqns soln} we find that:

$H_1$ has the Far amplitudes
\begin{align}
\begin{aligned}
C^f_1=C^f_2=&
	-{i\over\pi}\frac{2^{2 l+1} \Gamma\left(l+\frac{3}{2}\right)^2 (2\kappa)^{\frac{i\omega}{2\kappa}}}{(l+1) (l+2) \kappa ^{l+2} \omega ^{l+1} \left({i \omega }/{\kappa}\right)_{l+2}} \,, \\
C^f_3=C^f_4=&
	-{i\over\pi}\frac{3\cdot 2^{2 l+1} \Gamma\left(l+\frac{3}{2}\right)^2 (2\kappa)^{\frac{i\omega}{2\kappa}}}
	{l(l+1)^3 (l+2) \kappa ^{l+2} \omega ^{l} 
		\left({i \omega}/{\kappa}\right)_{l+2}} \,,
\end{aligned}
\end{align}

$H_2$ has the Far amplitudes
\begin{align}
\begin{aligned}
C^f_1=C^f_2=&
	{i\over\pi}\frac{3\cdot 2^{2 l+1}(3 l-1)  
	\Gamma \left(l+\frac{3}{2}\right)^2 (2\kappa)^{\frac{i\omega}{2\kappa}}}{l (l+1) (l+2) (2 l-1)(2 l+1) \kappa ^{l} \omega ^{l+1} \left({i \omega }/{\kappa }\right)_{l}} \,, \\
C^f_3=C^f_4=&
	{i\over\pi}\frac{2^{2 l+1}(25l-8)  
	\Gamma \left(l+\frac{3}{2}\right)^2 (2\kappa)^{\frac{i\omega}{2\kappa}}}{ l^2 (l+1)^3 (l+2) (2 l-1)(2 l+1) \kappa ^{l} \omega ^{l} \left({i \omega }/{\kappa }\right)_{l}} \,,
\end{aligned}
\end{align}

$H_3$ has the Near amplitudes
\begin{align}
\begin{aligned}
C^n_1=-\alpha C^n_2=&
	{i\over\pi}\frac{3\cdot 2^{2 l-1} (l-1) l (l+1) (2 l+1) 
	\Gamma \left(l+\frac{1}{2}\right)^2 (2\kappa)^{\frac{i\omega}{2\kappa}}}
	{\kappa^{l+2} \omega^{l-1} \left({i \omega }/{\kappa}\right)_{l+2}} \,, \\
C^n_3=-\beta C^n_4=&
	{i\over\pi}\frac{2^{2 l-1} (l-1) l (25 l+33) \Gamma\left(l+\frac{1}{2}\right)^2(2\kappa)^{\frac{i \omega }{2\kappa }}}{(2 l+3)\kappa^{l} \omega ^{l-1} \left({i \omega }/{\kappa}\right)_{l}} \,,
\end{aligned}
\end{align}

$H_4$ has the Near amplitudes
\begin{align}
\begin{aligned}
C^n_1=-\alpha C^n_2=&
	{i\over\pi}\frac{2^{2 l-1} (l-1) l^3 (l+1)^2 (2 l+1)  
	\Gamma \left(l+\frac{1}{2}\right)^2 (2\kappa)^{\frac{i \omega }{2\kappa }}}
	{\kappa^{l+2} \omega^{l}\left({i \omega }/{\kappa }\right)_{l+2}} \,, \\
C^n_3=-\beta C^n_4=&
	{i\over\pi}\frac{3\cdot 2^{2 l-1} (l-1) l^3 (l+1) (3 l+4) \Gamma \left(l+\frac{1}{2}\right)^2 (2\kappa)^{\frac{i \omega }{2\kappa }}}
	{(2 l+3) \kappa^{l} \omega^{l} \left({i \omega }/{\kappa }\right)_l} \,.
\end{aligned}
\end{align}

\section{Conclusion}\label{sec: conclusion}

In this paper we have analytically solved the coupled gravitational and electromagnetic perturbation equations for low energy modes propagating in the spacetime of a near-extreme Reissner-Nordstrom black hole. The computations herein generalize the ones obtained for extreme Reissner-Nordstrom in \cite{paper1} and as such they pave the way for transferring results from studies of near--$AdS_2$ holography to the realm of four-dimensional classical and quantum gravity in asymptotically flat spacetimes.
It is worth emphasizing that while the Near solution in this paper is diffeomorphic to the Near solution in \cite{paper1}, and the Static and Far solutions are identical, we have seen that the full Reissner-Nordstrom solution is markedly distinct. Holographically, the reason is that the mapping of Near solutions from Poincare to Rindler AdS is nontrivial: it is dual to a field theory mapping from zero to finite temperature, which when glued onto the Static and Far solutions gets propagated to null infinity.

From a purely classical gravitational physics viewpoint, this paper also provides a fresh look into the perturbation problem of generic Reissner-Nordstrom black holes. To date, this problem has been studied in terms of the well known reductions to two second-order radial equations first derived in the seventies (see \emph{e.g.} the monograph \cite{Chandrasekhar:1985kt}). In this paper, we have found a new reduction of the perturbation equations to a single fourth-order equation (Section \ref{sec: reduction of perturbation eqs}). Moreover, in the case of a black hole near extremality, we have found an analytic solution for low energy modes (Section \ref{sec: main}). Analytic solutions of Einstein's equations are among the rarest gems and they may inform even numerical studies, particularly near extremality where the numerics begin to struggle.

The results contained in this paper --and in \cite{paper1}-- may prove useful in analytic studies of a wide range of applications that hinge on the perturbative solution of Einstein's equations around extreme and near-extreme Reissner-Nordstrom black holes. This includes computations of radiation fluxes and absorption cross-sections \cite{Olson:1974nk, Crispino:2009zza}, quasinormal modes \cite{Onozawa:1995vu, Andersson:1996xw, Zimmerman:2015trm}, late-time tails \cite{Lucietti:2012xr, Sela:2016qau}, backreaction and self-force effects \cite{Isoyama:2011ea, Zimmerman:2012zu}, supergravity scattering amplitudes relations \cite{Okamura:1997ic}, and more.

\acknowledgements

This work is supported by NSF grant PHY-1504541


\appendix
\section{The master radial equation for Reissner-Nordstrom perturbations}\label{app: a's}

In Section \ref{sec: reduction of perturbation eqs} we performed a reduction of the general Reissner-Nordstrom perturbation equations to the fourth-order differential equation \eqref{Keqn}.
The coefficients $a_i(r)$ in Eq.~\eqref{Keqn} are given by:

\begin{dgroup*}
	
\begin{dmath*}
a_4(r)=r^4 (1+r)^4 (r+2 \kappa +r \kappa )^4 \left(\lambda ^2+4 (1+r)^2 (1+\kappa )^2 \omega ^2\right)^2
\end{dmath*}

\begin{dmath*}
a_3(r)=2 r^3 (1+r)^3 (r+2 \kappa +r \kappa )^3 \left(\lambda ^2+4 (1+r)^2 (1+\kappa )^2 \omega ^2\right) \left(\lambda ^2 \left(4 r+r^2+4 \kappa +2 r \kappa +r^2 \kappa
\right)+4 (1+r)^2 (1+\kappa )^2 \left(4 r-r^2+4 \kappa -2 r \kappa -r^2 \kappa \right) \omega ^2\right)
\end{dmath*}

\begin{dmath*}
a_2(r)=2 r^2 (1+r)^2 (r+2 \kappa +r \kappa )^2 \left[-\lambda ^4 \left(\lambda  r (1+r)^2 (1+\kappa ) (r+2 \kappa +r \kappa )-2 \left(3 r^2-2 r^3+6 r \kappa -6 r^2 \kappa -2 r^3 \kappa +2 \kappa ^2-6 r \kappa ^2-3 r^2 \kappa^2\right)\right)
+\lambda ^2 (1+r)^2 \left(\lambda ^2 (1+r)^4 (1+\kappa )^4-4 \lambda  r (1+r)^2 (1+\kappa )^3 (r+2 \kappa +r \kappa )+4 (1+\kappa)^2
\\
 \left(12 r^2-16 r^3-r^4+24 r \kappa -48 r^2 \kappa -20 r^3 \kappa -2 r^4 \kappa +8 \kappa ^2-40 r \kappa ^2-24 r^2 \kappa^2-4 r^3 \kappa ^2-r^4 \kappa^2\right)\right) \omega^2
+8 (1+r)^4 (1+\kappa )^4 \left(\lambda ^2 (1+r)^4 (1+\kappa)^2+2 \left(6 r^2-12 r^3+3 r^4+12 r \kappa -36r^2 \kappa +6 r^4 \kappa +4 \kappa ^2-28 r \kappa^2-2 r^2 \kappa ^2+12 r^3 \kappa ^2+3 r^4 \kappa ^2\right)\right) \omega^4
+16 (1+r)^{10} (1+\kappa )^8 \omega
^6\right]
\end{dmath*}

\begin{dmath*}
a_1(r)=2 r^2 (1+r) (r+2 \kappa +r \kappa )^2 \left[-\lambda ^4 \left(\lambda  (1+r)^2 (1+\kappa ) \left(2 r-r^2+2 \kappa -2 r \kappa -r^2 \kappa \right)+4 \left(3 r^2-2 r^3+6r \kappa -6 r^2 \kappa -2 r^3 \kappa +2 \kappa ^2-6 r \kappa ^2-3 r^2 \kappa ^2\right)\right)
+\lambda^2 (1+r)^2 (1+\kappa)^2 \left(\lambda ^2 (1+r)^4 (1+\kappa)^2-4 \lambda  (1+r)^2 (1+\kappa ) \left(2 r-5 r^2+2 \kappa -10 r \kappa -5 r^2 \kappa \right)-8 \left(12 r^2-16 r^3-r^4+24 r \kappa -48 r^2 \kappa -20 r^3 \kappa-2 r^4 \kappa +8 \kappa ^2-40 r \kappa ^2-24 r^2 \kappa ^2-4 r^3 \kappa^2-r^4 \kappa ^2\right)\right) \omega^2
-32 (1+r)^4 (1+\kappa )^4 
\left(6 r^2-12 r^3+3 r^4+12 r \kappa -36 r^2 \kappa +6 r^4 \kappa +4 \kappa ^2-28 r \kappa ^2-2 r^2 \kappa ^2+12 r^3 \kappa ^2+3 r^4 \kappa ^2\right) \omega^4
\\
-16 (1+r)^{10} (1+\kappa)^8 \omega^6\right]
\end{dmath*}

\begin{dmath*}
a_0(r)=\lambda ^4 r^2 (r+2 \kappa +r \kappa )^2 \left(\lambda ^2 (1+r)^4 (1+\kappa )^2-2 \lambda  (1-r) (1+r)^2 (1+\kappa ) (1-r-3 \kappa -r \kappa )+8 \left(3 r^2-2 r^3+6 r
\kappa -6 r^2 \kappa -2 r^3 \kappa +2 \kappa ^2-6 r \kappa ^2-3 r^2 \kappa ^2\right)\right)-2 \lambda ^2 (1+r)^2 (1+\kappa )^2 \left(\lambda ^3 r (1+r)^6 (1+\kappa
)^3 (r+2 \kappa +r \kappa )+2 \lambda ^2 (1+r)^4 (1+\kappa )^2 \left(r^3-r^4+3 r^2 \kappa -3 r^3 \kappa -2 r^4 \kappa -\kappa ^2-4 r^2 \kappa ^2-4 r^3 \kappa ^2-r^4
\kappa ^2\right)+4 \lambda  r^2 (1+r)^2 (1+\kappa ) (r+2 \kappa +r \kappa )^2 \left(4-r+5 r^2-5 \kappa +10 r \kappa +5 r^2 \kappa \right)-8 r^2 (r+2 \kappa +r
\kappa )^2 \left(12 r^2-16 r^3-r^4+24 r \kappa -48 r^2 \kappa -20 r^3 \kappa -2 r^4 \kappa +8 \kappa ^2-40 r \kappa ^2-24 r^2 \kappa ^2-4 r^3 \kappa ^2-r^4 \kappa
^2\right)\right) \omega ^2+(1+r)^4 (1+\kappa )^4 \left(\lambda ^4 (1+r)^8 (1+\kappa )^4-8 \lambda ^3 r (1+r)^6 (1+\kappa )^3 (r+2 \kappa +r \kappa )+8 \lambda ^2
(1+r)^4 (1+\kappa )^2 \left(7 r^4+28 r^3 \kappa +14 r^4 \kappa +4 \kappa ^2+8 r \kappa ^2+32 r^2 \kappa ^2+28 r^3 \kappa ^2+7 r^4 \kappa ^2\right)-96 \lambda  r^2
(1+r)^2 (1+\kappa ) (1-r-2 \kappa ) (r+2 \kappa +r \kappa )^2+64 r^2 (r+2 \kappa +r \kappa )^2 \left(6 r^2-12 r^3+3 r^4+12 r \kappa -36 r^2 \kappa +6 r^4 \kappa +4
\kappa ^2-28 r \kappa ^2-2 r^2 \kappa ^2+12 r^3 \kappa ^2+3 r^4 \kappa ^2\right)\right) \omega ^4+8 (1+r)^{10} (1+\kappa )^8 \left(\lambda ^2 (1+r)^4 (1+\kappa
)^2+4 \left(2 r^3+3 r^4+6 r^2 \kappa +14 r^3 \kappa +6 r^4 \kappa +2 \kappa ^2+8 r \kappa ^2+16 r^2 \kappa ^2+12 r^3 \kappa ^2+3 r^4 \kappa ^2\right)\right) \omega
^6+16 (1+r)^{16} (1+\kappa )^{12} \omega ^8
\end{dmath*}

\end{dgroup*}

\end{document}